\documentclass[twocolumn]{aastex631}

\received{\today}
\revised{\today}
\accepted{\today}

\submitjournal{ApJ}

\shorttitle{A flux rope eruption observed by SO \& PSP}
\shortauthors{Long et al.}

\newcommand{\cf}{{cf.}}
\newcommand{\corr}[1]{\textcolor{black}{#1}} % Modifications following referee report

\begin{document}

\title{The eruption of a magnetic flux rope observed by \emph{Solar Orbiter} and \emph{Parker Solar Probe}}

\correspondingauthor{David M.~Long}
\email{d.long@qub.ac.uk}

\author[0000-0003-3137-0277]{David M.~Long}
\affiliation{Astrophysics Research Centre, School of Mathematics and Physics, Queen’s University Belfast, University Road, Belfast, BT7 1NN, Northern Ireland, UK}

\author[0000-0002-0053-4876]{Lucie~M.~Green}
\affiliation{University College London, Mullard Space Science Laboratory, Holmbury St. Mary, Dorking, Surrey, RH5 6NT, UK}

\author[0000-0003-4168-590X]{Francesco Pecora}
\affiliation{Department of Physics and Astronomy, University of Delaware, Newark, DE 19716, USA}

\author[0000-0002-2189-9313]{David H.~Brooks}
\affiliation{College of Science, George Mason University, 4400 University Drive, Fairfax, VA 22030, USA}

\author[0000-0003-1483-4535]{Hanna Strecker}
\affiliation{Instituto de Astrofísica de Andalucía (CSIC), Apdo. de Correos 3004, E-18080 Granada, Spain}

\author[0000-0001-8829-1938]{David Orozco-Su\'{a}rez}
\affiliation{Instituto de Astrofísica de Andalucía (CSIC), Apdo. de Correos 3004, E-18080 Granada, Spain}

\author[0000-0002-6835-2390]{Laura A.~Hayes}
\affiliation{ESTEC, European Space Agency, Keplerlaan 1, PO Box 299, NL-2200 AG Noordwijk, The Netherlands}

\author[0000-0001-9992-8471]{Emma E.~Davies}
\affiliation{Austrian Space Weather Office, GeoSphere Austria, 8020 Graz, Austria}

\author[0000-0003-1516-5441]{Ute V.~Amerstorfer}
\affiliation{Austrian Space Weather Office, GeoSphere Austria, 8020 Graz, Austria}

\author[0000-0003-4105-7364]{Marilena Mierla}
\affiliation{Solar-Terrestrial Centre of Excellence -- SIDC, Royal Observatory of Belgium, Ringlaan 3, 1180 Brussels, Belgium}
\affiliation{Institute of Geodynamics of the Romanian Academy, 020032 Bucharest-37, Romania}

\author[0000-0002-3176-8704]{David Lario}
\affiliation{Heliophysics Science Division, NASA Goddard Space Flight Center, Greenbelt, MD 20771, USA}

\author[0000-0003-4052-9462]{David Berghmans}
\affiliation{Solar-Terrestrial Centre of Excellence -- SIDC, Royal Observatory of Belgium, Ringlaan 3, 1180 Brussels, Belgium}

\author[0000-0002-2542-9810]{Andrei N. Zhukov}
\affiliation{Solar-Terrestrial Centre of Excellence -- SIDC, Royal Observatory of Belgium, Ringlaan 3, 1180 Brussels, Belgium}
\affiliation{Skobeltsyn Institute of Nuclear Physics, Moscow State University, 119991 Moscow, Russia}

\author[0000-0002-2559-2669]{Hannah T.~R\"{u}disser}
\affiliation{Austrian Space Weather Office, GeoSphere Austria, 8020 Graz, Austria}

%%%%%%%%%%%%%%%%%%%%%%%%%%%%%%%%%%%%%%%%%%%%%%%%%%%
%%% Abstract 
\begin{abstract}
Magnetic flux ropes are a key component of coronal mass ejections, forming the core of these eruptive phenomena. However, \corr{determining whether a flux rope is present prior to eruption onset and, if so, the rope's handedness and the number of turns that any helical field lines make} is difficult without magnetic field modelling or in-situ detection of the flux rope. We present two distinct observations of plasma flows along a \corr{filament channel} on 4 and 5 September 2022 \corr{made} using the \emph{Solar Orbiter} spacecraft. Each plasma flow exhibited \corr{helical motions in a right-handed sense as the plasma moved} from the source active region across the solar disk to the quiet Sun, suggesting \corr{that the magnetic configuration of the filament channel contains a flux rope with} positive \corr{chirality and at least one turn}. The length and velocity of the plasma flow increased from the first to the second observation, suggesting evolution of the flux rope, with the flux rope subsequently erupting within $\sim$5~hours of the second plasma flow. The erupting flux rope then passed over the \emph{Parker Solar Probe} spacecraft during its Encounter 13, enabling \emph{in-situ} diagnostics of the structure. Although complex and consistent with the flux rope erupting from underneath the heliospheric current sheet, the \emph{in-situ} measurements support the inference of a right-handed flux rope from remote-sensing observations. These observations provide a unique insight into the eruption and evolution of a magnetic flux rope near the Sun.
\end{abstract}

%%%%%%%%%%%%%%%%%%%%%%%%%%%%%%%%%%%%%%%%%%%%%%%%%%%
%% Keywords
%

%-------------------------------------------------

%%%%%%%%%%%%%%%%%%%%%%%%%%%%%%%%%%%%%%%%%%%%%%%%%%%
%% Sections
%
\section{Introduction}\label{s:intro} 

As the coronal magnetic field is rooted in the constantly moving photosphere, it is continuously twisting and shearing across a range of scales. Through a process of small-scale magnetic flux cancellation, this can lead to the development of twisted/sheared magnetic field structures called magnetic flux ropes \citep[cf.][]{vanball:1989}. These features (also called filament channels) are commonplace on the Sun, appearing along polarity inversion lines as elongated low emission structures in extreme ultraviolet (EUV) remote sensing observations of the solar corona. Lower down in the chromosphere, they are characterised by fibrils that are also highly aligned to the polarity inversion line \citep[cf.][]{Babcock:1955,Martin:1988}. 

As magnetic structures, flux ropes are difficult to identify in the optically thin and high temperature plasma of the low-$\beta$ solar corona. However they are easier to identify using \emph{in-situ} measurements as coherent rotating magnetic field \citep[cf.][]{Gosling:1990} following eruption into interplanetary space. Given this difficulty in pre-eruption identification and their strong relationship with coronal mass ejections (CMEs), there has been discussion as to whether they are formed prior to or during a CME eruption \citep[see, e.g., the reviews by][]{Forbes:2000,Forbes:2006}. \corr{Observational determination of whether or not a flux rope is present in the pre-eruptive magnetic field acts as a discriminator between CME models that do \cite[e.g.][]{Forbes:1991,Toeroek:2005, Kliem:2006} and do not \cite[e.g.][]{Antiochos:1999,Moore:2001} require this structure for an eruption to occur.}

Magnetic flux ropes forming prior to the CME eruption \corr{can} be produced by magnetic reconnection in the photosphere/chromosphere associated with flux cancellation, as described by \citet{vanball:1989}. Flux ropes formed in this way would be expected to exhibit at least one full turn of an off-axial poloidal magnetic field twisted around an axial toroidal magnetic field, and could support filamentary material in concave-up \corr{sections of the} magnetic field in the underside of the flux rope. \corr{In contrast, a flux rope formed due to flaring reconnection would originally take the form of a sheared arcade of magnetic field during its pre-eruptive phase}. Models describing flux rope formation in this way include magnetic reconnection within a sheared arcade \citep[tether-cutting reconnection;][]{Moore:2001} or magnetic reconnection in an inflating sheared arcade \citep[the second phase of reconnection in the breakout model;][]{Antiochos:1999}. For more details on the origin and initial evolution of coronal mass ejections, see the recent reviews by \citet{Green:2019,Patsourakos:2020}, and references therein. Regardless of the formation mechanism, the flux rope would then be identifiable as a dark cavity within the erupting CME \corr{when viewed in coronagraph images} \citep[e.g.,][]{Dere:1999}, and exhibit the same coherent rotating magnetic field \emph{in-situ}.

Despite this ambiguity, there is a growing body of observational evidence for the presence of a magnetic flux rope in the solar atmosphere prior to the CME eruption. Notwithstanding the difficulty in directly determining the presence of a magnetic flux rope via measurement of the coronal magnetic field, they can be identified via extrapolation of the photospheric magnetic field \citep[e.g.,][]{James:2018,Yardley:2019}, or via certain observational signatures using coronal observations. In particular, EUV and soft X-ray emission structures that exhibit an S-shape \citep[e.g.,][]{Green:2007,Green:2009}, so-called hot flux rope features that are formed via reconnection in the corona 
%during confined flares 
\citep[e.g.,][]{Zhang:2012,Patsourakos:2013,James:2017}, and low-density coronal cavities observed at the limb in EUV or white light coronagraph data \citep[e.g.,][]{Gibson:2006,Sarkar:2019,Gibson:2015} can all be taken as evidence of the existence of a magnetic flux rope in the solar corona.

The magnetic configuration of flux ropes also means that they can support relatively dense and cool plasma in the concave-up magnetic field in their underside \citep[e.g.,][]{Aulanier:1998,Demoulin:1998}. Unlike the magnetic field, this plasma can then be observed in absorption on-disk as a dark filament or in emission above the limb as a bright prominence \citep[for more details, see the review by][]{Parenti:2014}. As filaments are features that are readily identifiable using ground-based H-$\alpha$ observations, they have long been identified and characterised based on their location of formation \citep[cf.][]{Gibson:2018,Mackay:2012}. Filaments tend to form either within the strong magnetic field structures of active regions, (where they are called “active region filaments”), between active regions (“intermediate filaments”), or wholly in the quiet Sun (“quiescent filaments”). In each case, the formation mechanism and \corr{timescale may vary} significantly, indicating that filaments can exhibit a range of plasma and magnetic field parameters. 

\corr{Filaments form within filament channels, the magnetic field configuration of which is currently hard to determine.}
As with other long-lived coronal structures such as coronal holes, filaments can \corr{be present} on the Sun for extended periods of time. 
However, they can become unstable as a result of e.g., nearby flux emergence \citep[cf.][]{Chen:2000,Feynman:1995}, or mass unloading of their filamentary plasma \citep[cf.][]{Jenkins:2018,Seaton:2011}. When this happens, \corr{filaments and the filament channel} can erupt into the heliosphere as CMEs \citep[see e.g., the recent review by][]{Patsourakos:2020}. As such, understanding the formation and evolution of filaments as well as identifying the triggers which lead to their eruption is of vital importance for space weather research and continues to be the source of detailed investigation \citep[e.g.][]{Liu:2020}.

Here we describe two distinct observations of plasma flows along a \corr{filament channel} which exhibited clear right-handed evolution \corr{and which appeared to trace out a flux rope configuration}. This flux rope subsequently erupted and was detected \emph{in-situ} within 14~R$_{\odot}$ of the Sun, with the \emph{in-situ} measurements confirming the flux rope properties inferred using remote-sensing observations. Section~\ref{s:obs} describes the observational data-sets used in this work. Section~\ref{s:res} outlines the different observational results, including the evolution of the magnetic field and associated plasma flows, eruption of the flux rope, and its subsequent \emph{in-situ} detection. Section~\ref{s:disc} then discusses these results, before some conclusions are drawn in Section~\ref{s:conc}.

\begin{figure*}[!t]
    \centering
    \includegraphics[width=0.97\textwidth]{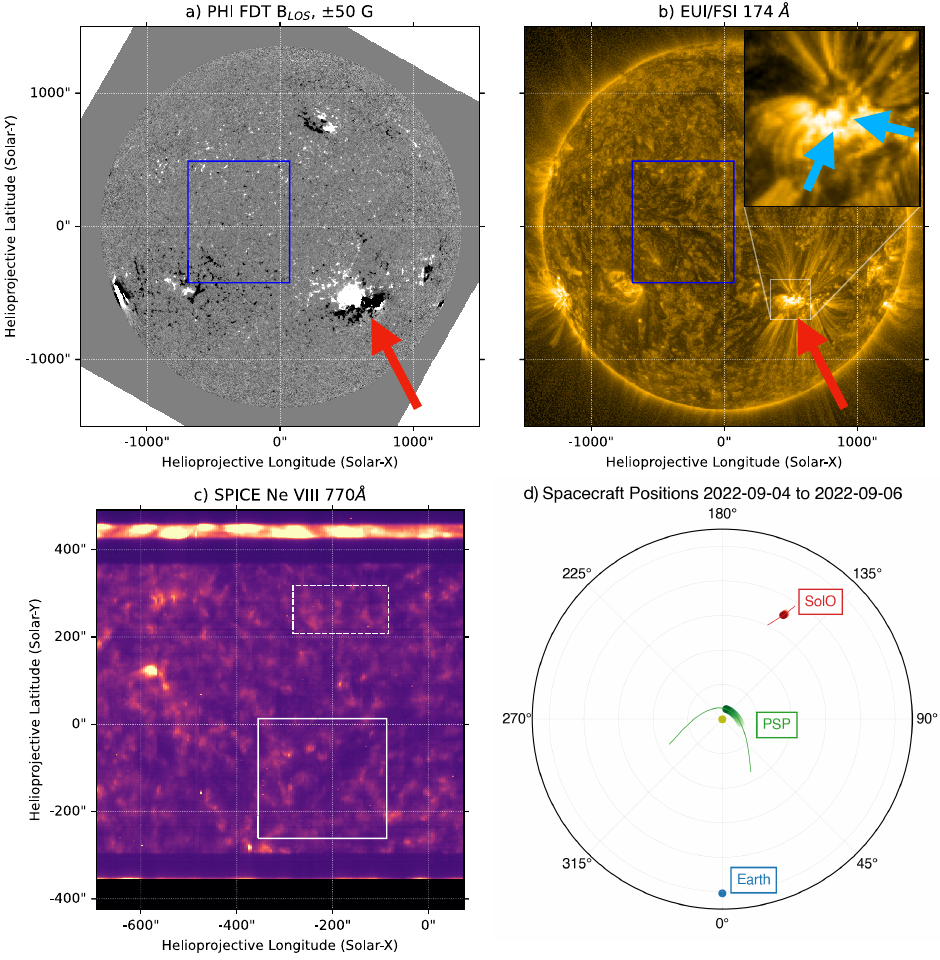}
    \caption{The Sun at 04:00~UT on 5~September~2022 as seen by the different remote sensing instruments onboard \emph{Solar Orbiter}. Panel~a shows the line of sight magnetic field observed by PHI/FDT, saturated at $\pm$50~G, panel~b shows the low corona observed using the EUI/FSI 174~\AA\ passband and processed using the Multiscale Gaussian Normalisation (MGN) technique of \citet{Morgan:2014}, with the inset showing a close-up of the source active region. Blue arrows show the active region filament above the internal polarity inversion line. Panel~c shows the SPICE Ne~VIII raster summed across the spectral window, with the solid and dashed squares corresponding to the blue and red spectra respectively shown in Figure~\ref{fig:spectrum}. The blue squares in panels a \& b show the SPICE field of view, with the red arrow indicating the source active region. Panel~d shows the location of \emph{Solar Orbiter} and \emph{Parker Solar Probe} during the period studied here.}
    \label{fig:SolO_context}
\end{figure*}

\section{Observations \& Data Analysis}\label{s:obs} 

On 4~September~2022, the \emph{Solar Orbiter} spacecraft \citep{Garcia:2021,Mueller:2020} was undergoing a gravity assist maneuver (GAM) at Venus, 0.716~astronomical units (au) from the Sun and $\sim$150$^{\circ}$ ahead of the Earth on approach to its second science perihelion (see Figure~\ref{fig:SolO_context}d). At the same time, \emph{Parker Solar Probe} \citep[PSP;][]{Fox:2016}, \corr{which} was undertaking Encounter 13, reached a perihelion of 0.062~au on 6~September. The positioning of both PSP and \emph{Solar Orbiter} on the far side of the Sun to the Earth (cf. Figure~\ref{fig:SolO_context}d) meant that \emph{Solar Orbiter} was perfectly positioned to provide remote sensing observations of the Sun in support of PSP Encounter~13.

To this end, following the Venus GAM, the Full Sun Imager (FSI), part of the Extreme Ultraviolet Imager \citep[EUI;][]{Rochus:2020} began taking synoptic observations in the 174~\AA\ and 304~\AA\ passbands from 03:30~UT on 4~September~2022 at a cadence of 10~minutes and 15~minutes respectively. EUI/FSI has a pixel scale of 4.44''/pixel and a field of view of (228')$^2$, enabling it to observe the full solar disk and corona simultaneously even at perihelion. The observations described here were analysed using Level-2 EUI FSI data from Data Release 6\footnote{EUI Data Release 6; doi:\href{https://doi.org/10.24414/z818-4163}{10.24414/z818-4163}}. The Level-2 data provided by the EUI team have already been fully calibrated from the Level-1 data using \mbox{\emph{euiprep.py}} which accounts for instrument deviations and spacecraft pointing instabilities \citep[see, e.g.,][]{Kraaikamp:2023}. Note that although the phenomenon was observed using both the 174 and 304~\AA\ passbands, the 174~\AA\ passband was used for this analysis due to the higher cadence and better signal-to-noise of the observations.

Remote sensing \emph{Solar Orbiter} support of PSP Encounter 13 also included synoptic spectroscopic observations made by the Spectral Investigation of the Coronal Environment \citep[SPICE;][]{SPICE:2020}. SPICE has a spatial and spectral resolution of 2'' and 0.04~nm respectively, and began taking a series of synoptic rasters with a 12~hour cadence from 03:34~UT on 5~September~2022. The SPICE data used here are calibrated level~2 data prepared and released as part of Data Release 3\footnote{SPICE Data Release 3; doi:\href{https://doi.org/10.48326/idoc.medoc.spice.3.0}{10.48326/idoc.medoc.spice.3.0}}. 

There were also synoptic observations of the photospheric magnetic field provided by the Full Disk Telescope (FDT) of the Polarimetric and Helioseismic Imager \citep[PHI;][]{Solanki:2020}. PHI/FDT has a resolution of 3".75/pixel and a field of view of 2$^{\circ}$, and began taking synoptic observations with a 3~hour cadence from 01:00~UT on 5~September~2022. The data used for this analysis was processed with the PHI/FDT on-ground pipeline which includes some changes in the reduction and processing of the data in comparison to the on-board pipeline \citep{Albert:2020}, which are in particular the application of a fringes and a ghost correction. The PHI/FDT data shown here were then rotated to put solar north up to match the observations from EUI and SPICE.

Following its eruption from the Sun, the flux rope was detected and analysed \emph{in-situ} using magnetic field measurements made by the FIELDS instrument \citep{Bale:2016} onboard \emph{Parker Solar Probe}. Fully calibrated level~2 data released as part of Data Release 14\footnote{FIELDS data: \href{http://research.ssl.berkeley.edu/data/psp/data/sci/fields/l2/}{https://fields.ssl.berkeley.edu/data/}} were used here. 

\section{Results}\label{s:res} 

\begin{figure*}[!t]
    \centering
    \includegraphics[width=0.98\textwidth]{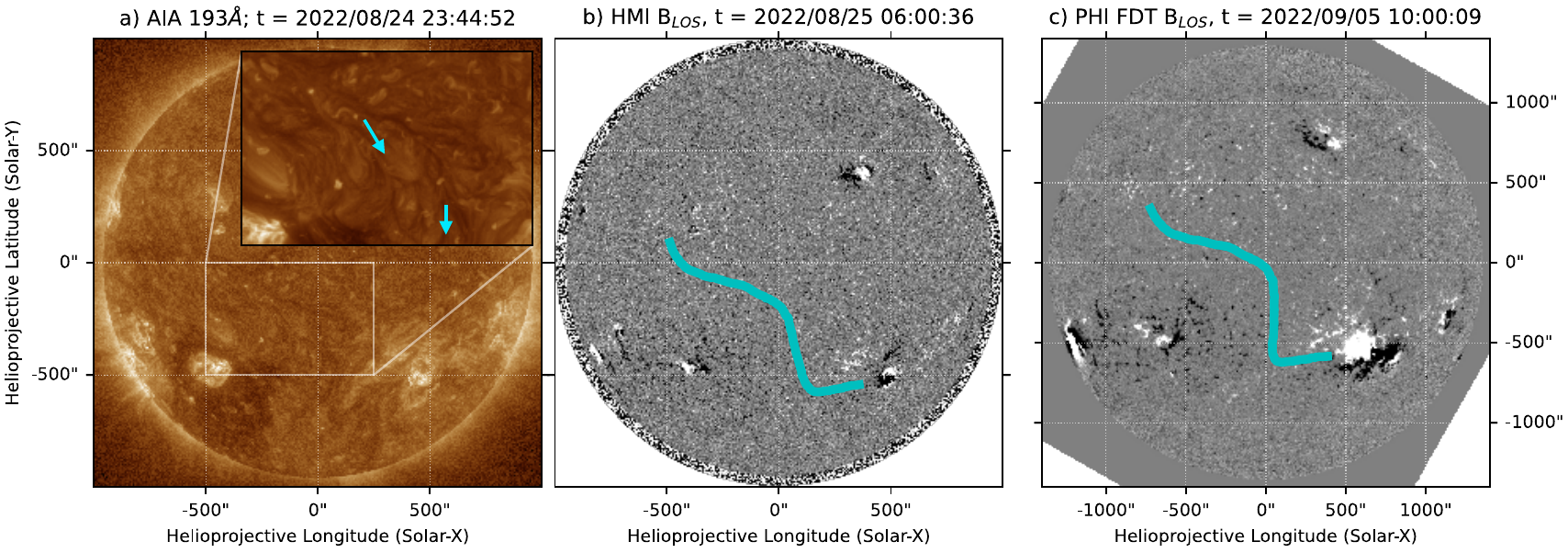}
    \caption{The filament channel seen using the 193~\AA\ passband on SDO/AIA on 24~August (panel~a), and line-of-sight photospheric magnetic field observed by \emph{SDO}/HMI on 25~August (panel~b) and \emph{Solar Orbiter}/PHI on 5~September (panel~c). The cyan arrows in panel~a show the direction of the plumes associated with the filament channel, indicating a sinistral configuration (see text for details). The cyan lines in panels~b \& c show the path of the plasma flow identified using EUI/FSI on 5~September as shown in Figure~\ref{fig:brightening}. Both magnetograms have been saturated at $\pm$50~G.}
    \label{fig:magnetograms}
\end{figure*}

The different phenomena described here occurred near disk centre as observed by \emph{Solar Orbiter} on 4 \& 5~September~2022 as seen in Figure~\ref{fig:SolO_context}. This figure shows the Sun as observed by PHI/FDT (panel~a), EUI/FSI (panel~b), and SPICE (panel~c, with the field of view on the full disk indicated by the blue box in panels a \& b) at $\sim$04:00~UT on 5~September. Although it is difficult to identify any features in the SPICE observations, a lower intensity region can be seen in the EUI image shown in Figure~\ref{fig:SolO_context}b from bottom right to top left of the blue box corresponding to the SPICE field of view. There also appears to be a change in magnetic polarity in the same region, with a separation between the mainly black, negative magnetic field in the bottom left to the mainly white, positive magnetic field in the top right of Figure~\ref{fig:SolO_context}a. This is suggestive of a magnetic inversion line, and combined with the low intensity seen in EUV observations, is consistent with the existence of a filament channel. The data presented later in this study support the interpretation that the magnetic field configuration of the filament channel is that of a flux rope.

\subsection{Magnetic field configuration and evolution}\label{ss:formation}

Although the main body of the filament channel is located over a quiet Sun region, its western end is rooted in the strong magnetic field of NOAA active region (AR) 13088 (indicated by the red arrow in panels a \& b of Figure~\ref{fig:SolO_context}). AR~13088 was formed by the rapid emergence of flux into the pre-existing negative polarity of a very dispersed and spotless bipolar active region (\corr{likely} corresponding to NOAA AR~13066 in the previous rotation) on 24~August~2022. Although both the dispersed bipolar region and AR~13088 had negative polarity leading magnetic field, the emerging flux initially emerged oriented primarily north-south as observed by HMI but was then observed by PHI to have negative leading and positive following polarity, indicating some rotation of the magnetic field early in its evolution. 

The filament channel was clearly observed on 24~August~2022 enabling an analysis using AIA data. No dense cool plasma can be identified in the filament channel which could be used to probe the magnetic field configuration. However, cellular features in the corona observed in emission at EUV wavelengths can be used to infer the chirality of the magnetic field of a filament channel \citep{Sheeley:2013}, in a way that is analogous to using chromospheric fibrils detected in H$\alpha$ imaging data \citep{Martin:1994}. The cellular features are known as cellular plumes (as they are tapered at one end) and are most readily seen in the AIA 193~\AA\ waveband (cf. Figure~\ref{fig:magnetograms}a) that images light from plasma at a temperature of $\sim$1.2MK \citep{Sheeley:2013}. \corr{Previous studies have shown that these} plumes exhibit a systematic orientation that is close to horizontal and \corr{they} are interpreted as being influenced by the axial magnetic field along the polarity inversion line of the filament channel \corr{\citep[e.g.,][]{Sheeley:2013,Su:2012}}. When viewed from the positive polarity side of the polarity inversion line, cellular plumes rooted in the positive field that are tapered on their right (left) hand-side and fan out to the left (right) indicate the presence of a sinistral \corr{(dextral)} structure \citep{Sheeley:2013}. \corr{Plumes that are rooted in the negative polarity field point in the opposite direction to their counterparts in the positive polarity side of the inversion line. In addition, in a sinistral (dextral) filament channel the axial field points to the left (right), again when viewed from the positive polarity side of the inversion line.} The configuration present in the filament channel studied here is sinistral\corr{, as indicated by the orientation of the plumes observed to be rooted in the negative polarity field of the filament channel and seen in the AIA 193~\AA\ waveband images (see the cyan arrows in the inset of Figure~\ref{fig:magnetograms}a). This indicates that the axial field points from right to left when viewed from the positive polarity side of the inversion line (or from north to south when viewed along the filament channel from the spacecraft perspective).} In a flux rope interpretation a sinistral structure indicates a right-handed twist in the magnetic field \citep{Chae:2000}.

AR~13088 rotated over the western limb as seen from the Earth perspective on 29~August. Given the longitudinal offset between Earth and \emph{Solar Orbiter}, AR~13088 appeared on-disk as seen by \emph{Solar Orbiter} on 30~August, and was well observed by the different remote-sensing instruments prior to their switch-off in advance of the \corr{\emph{Solar Orbiter}} Venus GAM on 3~September. The remote-sensing instruments onboard \emph{Solar Orbiter} began being switched on again from 4~September in preparation for PSP Encounter 13. 

The growth of AR~13088 can be seen by the increase in its area in the time between when the active region was observed by \emph{SDO}/HMI on 25~August and \emph{Solar Orbiter}/PHI on 5~September as shown in Figure~\ref{fig:magnetograms}. The \emph{Solar Orbiter}/PHI data (Figure~\ref{fig:magnetograms}b) show that the negative and positive polarities of the active region have rotated with respect to each other and also butted up against each other. Under these conditions it is likely that flux cancellation is taking place along the internal polarity inversion line of the active region, producing the small active region filament identified by the blue arrows in the subregion of Figure~\ref{fig:SolO_context}b. The cyan line in Figure~\ref{fig:magnetograms} corresponds to the path of the plasma flow observed by EUI/FSI on 5~September (see Section~\ref{ss:brightening} for more details). There is a separation between the dominantly negative (positive) magnetic field as shown in black (white) on the left (right) of this \corr{cyan} line. This indicates that the plasma flow observed by EUI/FSI flowed along a magnetic structure that ran along a quiet sun magnetic inversion line. Note that the regions on disk are slightly shifted in latitude as viewed from SDO and \emph{Solar Orbiter} due to the differing spacecraft positions. SDO was 7$^{\circ}$ north of the equator as determined in the Heliographic Carrington coordinate system, whereas \emph{Solar Orbiter} was 3$^{\circ}$ south of the equator.

\begin{figure*}[!t]
    \centering
    \begin{interactive}{animation}{context_animation.mp4}
    \includegraphics[width=0.95\textwidth]{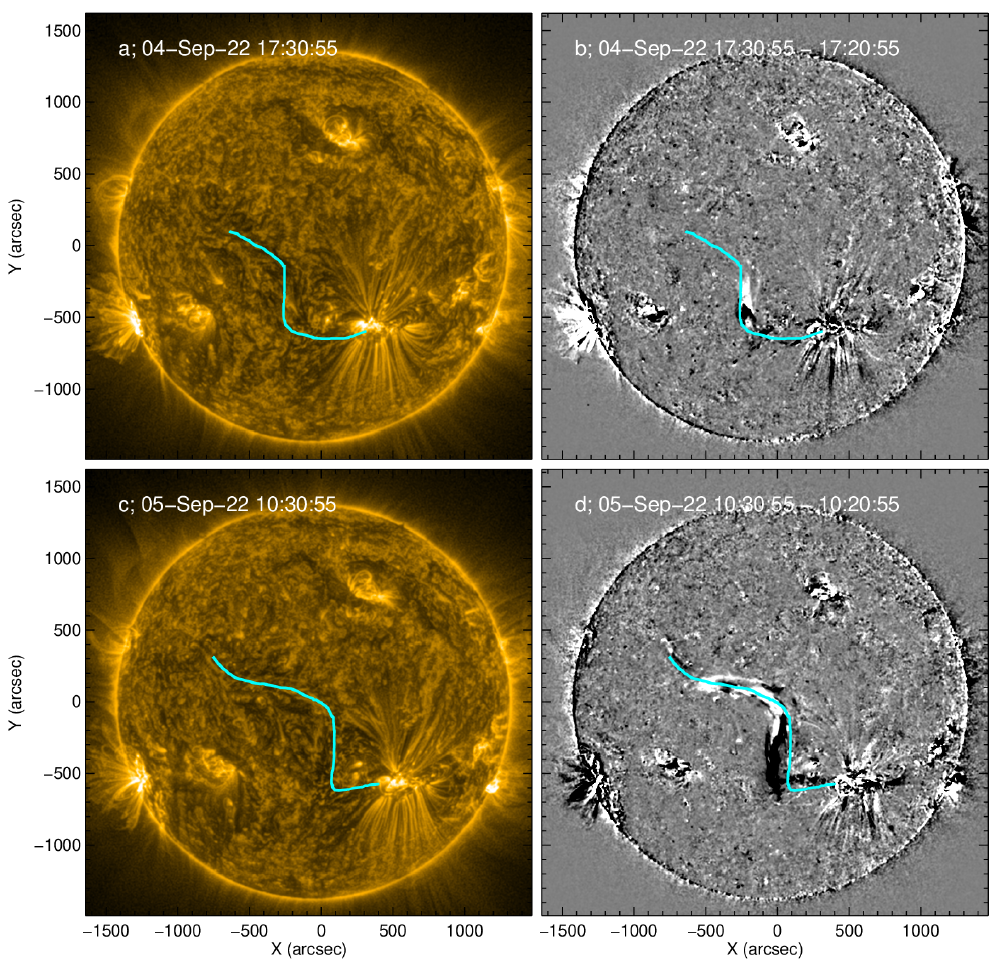}
    \end{interactive}
    \caption{The plasma flow along the filament channel observed on 4 (top row) and 5~September~2022 (bottom row). Left column shows the 174~\AA\ intensity image, and right column shows running difference images (produced by subtracting a preceding image from each subsequent image). The cyan line in each panel shows the spine of the filament channel used to estimate the velocity of the plasma flow in Figure~\ref{fig:kinematics}. An animated version of this figure is available as context\_animation.mp4, with a duration of 2~s which shows the temporal evolution of the plasma flow on 4 \& 5 September observed by Solar Orbiter EUI using intensity and running difference images.}
    \label{fig:brightening}
\end{figure*}

It should be noted that there is no clear evidence of any filamentary material along the region denoted by the cyan line between the emergence of the active region on 24~August and the first plasma flow observed by EUI/FSI on 4~September. However, small-scale flux emergence and cancellation can be seen in the HMI magnetograms due to its high cadence (PHI had a cadence of $\sim$3~hours at this time), suggesting flux rope development as predicted by \citet{vanball:1989}.

\begin{figure*}[!t]
    \centering
    \includegraphics[width=\textwidth]{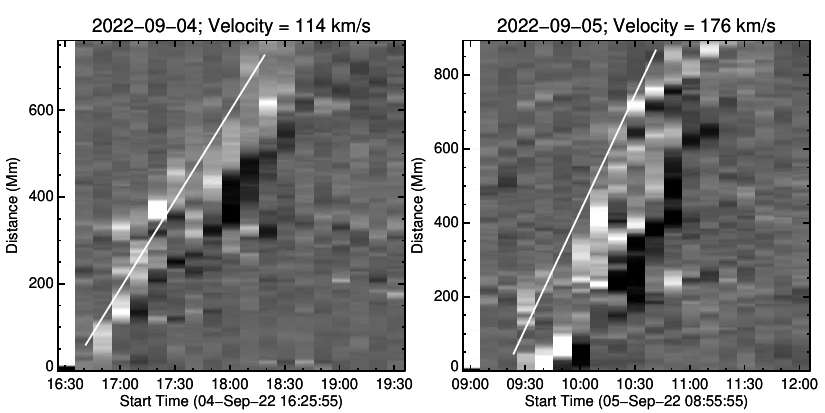}
    \caption{The temporal evolution of the plasma flow along the filament channel on 4 (left) and 5~September~2022 (right). In each case, the distance-time evolution of the plasma has been fitted using a linear fit (indicated by the white line) to derive the velocity (given in the title).}
    \label{fig:kinematics}
\end{figure*}

\subsection{Plasma diagnostics}\label{ss:brightening}

Approximately 14~hours after the switch-on of EUI/FSI in support of PSP Encounter 13, at $\sim$17:30~UT on 4~September, a plasma flow was observed originating at the active region at the bottom end of the cyan line in Figure~\ref{fig:brightening}a which then propagated along a snaking channel towards the quiet Sun. This phenomenon is best observed using the running difference image shown in Figure~\ref{fig:brightening}b and the associated animation. Within 16~hours of this plasma flow, at $\sim$10:15~UT on 5~September, a comparable plasma flow was observed again originating at the active region at the bottom end of the cyan line shown in Figure~\ref{fig:brightening}c, and propagating towards the quiet Sun (see, e.g., Figure~\ref{fig:brightening}d). The channel erupted $\sim$5~hours following the second plasma flow to produce a CME and a large associated flare in the origin active region. 

As shown in Figure~\ref{fig:brightening} and the associated animation, the two plasma flows observed on 4 and 5~September~2022 were qualitatively comparable, despite the $\sim$16~hour time difference between the phenomena. However, it is clear from the blue lines used to illustrate the flux rope channel in panels a \& c of Figure~\ref{fig:brightening} that both the length and large-scale writhe of the flux rope have increased in this 16~hour period. This suggests that the structure may have been becoming more unstable prior to its eruption. 

To quantify this progression towards the eruption, we first manually identified the spine of the flux rope as shown by the cyan lines in Figure~\ref{fig:brightening} by examining the full evolution of the plasma flow for each event. For both events, the data along this path were then used to produce distance-time stack plots as shown in Figure~\ref{fig:kinematics}. The leading edge of the bright front was then manually identified and fitted using a linear model to estimate the velocity along this path. This produced a plasma flow velocity of $\sim$114~km~s$^{-1}$ on 4~September, and of $\sim$176~km~s$^{-1}$ on 5~September, albeit over a longer path length (note the different y-axis range in the two panels of Fig.~\ref{fig:kinematics}).

\begin{figure*}[!t]
    \centering
    \includegraphics[width=\textwidth]{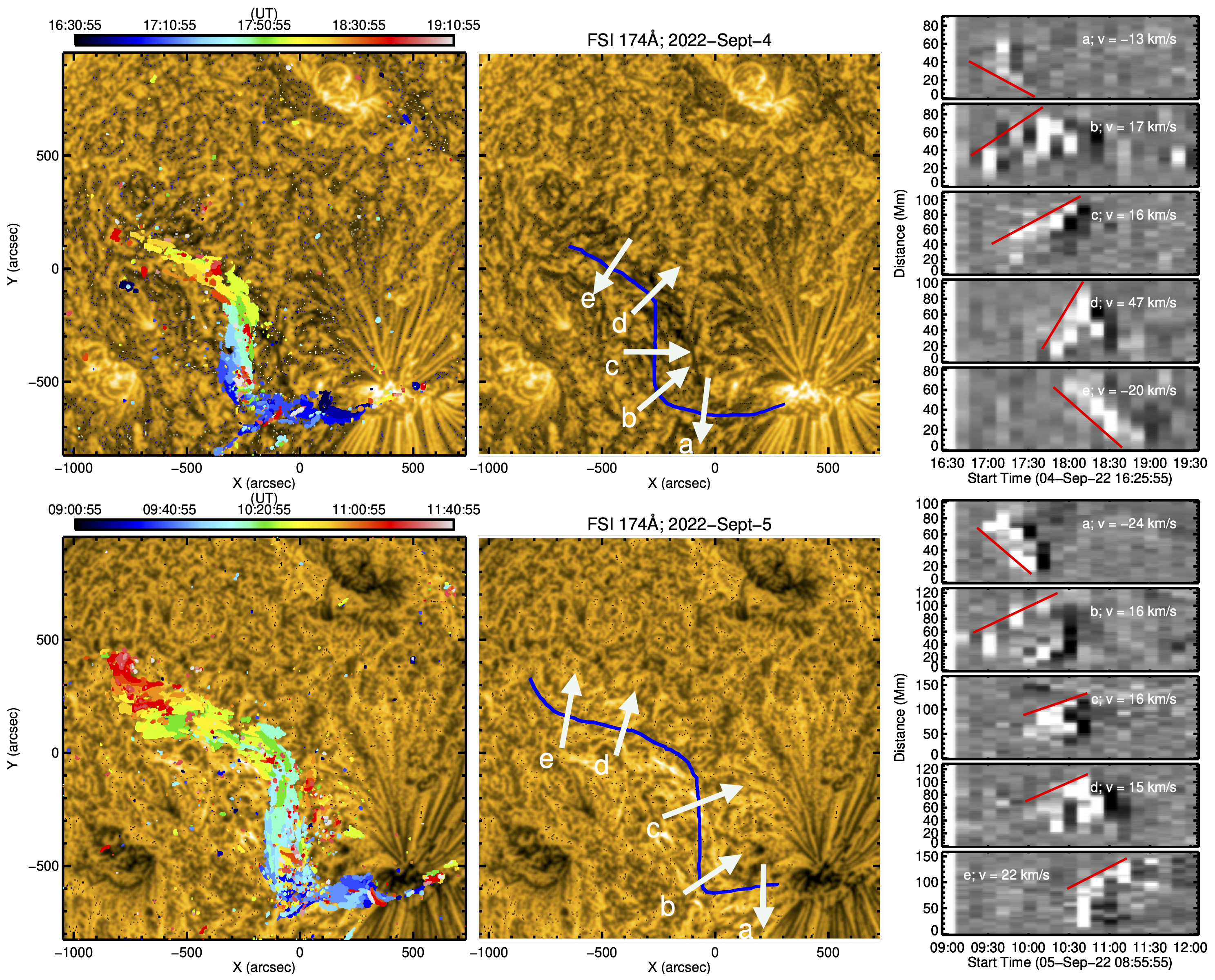}
    \caption{The small-scale evolution of the plasma flow along the observed flux rope structure on both 4 (top) and 5 (bottom) September~2022. Left column shows the plasma evolution derived by recording pixels with a running ratio value greater than 1.3 at each time step. Middle column shows the locations of the cuts taken across the blue spine of the flux rope structure corresponding to the distance-time plots in the right column. The arrows indicate the orientation of the plasma flow corresponding to the red linear fits in the distance-time plots.}
    \label{fig:profiles}
\end{figure*}

Although a line along the spine of the flux rope structure was used to estimate the bulk propagation velocity of the brightening along the identified path in Figure~\ref{fig:kinematics}, it is clear from Figure~\ref{fig:brightening} and the associated animation that the feature is quite broad for both events and has a distinctly different evolutionary pattern on either side of the spine. A series of running ratio images, produced by dividing each image by the previous image, were used to investigate this further. The pixels showing an intensity greater than 1.3 times the intensity of the previous time step were recorded for each time step, with these pixels then coloured according to the image time as shown in the left column of Figure~\ref{fig:profiles}. Using this approach, it is possible to identify a slight side-to-side variation across the feature from one end to the other. 

This apparent lateral motion was further investigated by taking a series of cuts across the feature as shown in the middle column of Figure~\ref{fig:profiles}. The intensity along each cut was then plotted with time as shown in the right column of Figure~\ref{fig:profiles}, with a linear fit (as illustrated by the red line) then used in each case to identify any left-to-right motion. The arrows in the middle column of Figure~\ref{fig:profiles} show the resulting propagation across the feature, with a positive (negative) linear fit to the distance-time plot corresponding to a left-to-right (right-to-left) motion.

The observed evolutionary behaviour is consistent with a plasma flow originating from the active region and propagating along the structure towards the quiet Sun with a right-handed motion about the central axis. On 4~September, the plasma appears to complete one turn, as evidenced by the reverse motion seen in cuts a and e, with each of the cuts exhibiting relatively low velocities. The exception to this is cut~d, which exhibits a high velocity, but is also co-spatial with a strong kink in the structure. This suggests a steep gradient in the middle of the three-dimensional structure, although the reversed direction of the velocity in cut~e suggests possible pooling of the plasma near the footpoint. These observations are consistent with plasma flow initially driven by energy release in the origin active region along a relatively stable structure towards the quiet Sun.

\begin{figure}[!t]
    \centering
    \includegraphics[width=0.47\textwidth]{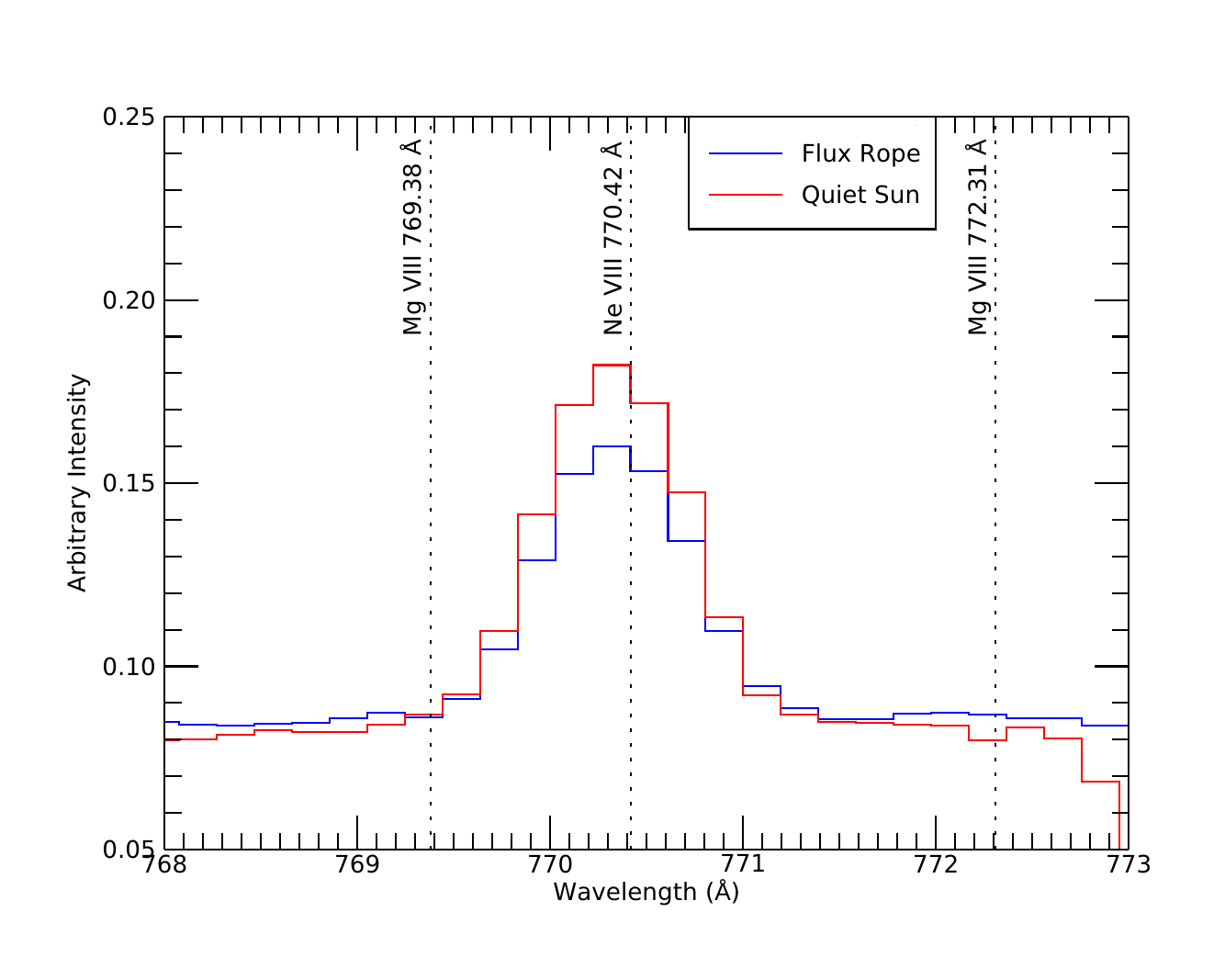}
    \caption{SPICE spectra of the flux rope (blue) and a control quiet Sun region (red). These spectra are averaged over the boxed areas shown in Figure \ref{fig:SolO_context}. The spectral positions of several Ne~VIII and Mg~VIII lines are indicated. The Mg~VIII lines should be bright relative to the Ne~VIII lines when there is a strong FIP effect in operation. }
    \label{fig:spectrum}
\end{figure}

The second plasma flow on 5~September shown in the bottom row of Figure~\ref{fig:profiles} again displays a twisted motion, in this case along a longer structure and with a much more pronounced ``elbow'' also observable close to the origin active region. The cuts across the spine of the structure show similar behaviour to that on 4~September, indicating a right-handed motion. However, the lateral flow across the structure has reversed direction in cut e, suggesting plasma draining down the leg of the structure towards the quiet Sun. This behaviour is consistent with increased plasma draining towards the quiet Sun as the structure becomes more unstable and rises slowly between the plasma flows observed on 4 and 5~September.

In addition to the FSI observations of the plasma flows, the SPICE spectrometer took a series of synoptic rasters of the region containing the flux rope (as noted in Section ~\ref{s:obs}, with the field of view shown in Figure~\ref{fig:SolO_context}a \& b). Previous work has shown that plasma composition can be used to determine where in the solar atmosphere a magnetic flux rope has formed, with \citet{Baker:2022} finding photospheric plasma composition within a flux rope which they interpreted as evidence for flux rope formation via magnetic flux cancellation in the photosphere. Here, we attempted to estimate the plasma composition in two distinct regions observed by SPICE. One region was chosen to be within the flux rope close to the origin active region AR~13088, with the other region chosen to be quiet Sun to the north (see boxes in Figure~\ref{fig:SolO_context}c). The spectra from these regions are shown in Figure~\ref{fig:spectrum}, with the flux rope spectrum plotted in blue (corresponding to the solid box in Figure~\ref{fig:SolO_context}c) and the quiet Sun spectrum plotted in red (corresponding to the dashed box in Figure~\ref{fig:SolO_context}c). It is clear that while the Ne~VIII lines can be observed in both cases, with the quiet Sun intensity higher than that of the flux rope, the Mg~VIII lines are practically nonexistent in both regions. If there was a strong first ionisation potential (FIP) effect in either region (indicating a strong FIP bias), we would expect the Mg~VIII lines to be bright relative to the Ne~VIII lines \citep{Brooks:2022}. This is clearly not the case in either region, and can be interpreted as showing that the composition is the same in both regions. This suggests that the flux rope has a composition that is similar to the quiet Sun\corr{. As there is no strong FIP effect, this implies photospheric plasma, consistent with observations by \citet{Lanzafame:2005} of photospheric abundances in the quiet corona, and matching the previous observations by \citet{Baker:2022}}.

\begin{figure*}[!t]
    \centering
    \includegraphics[width=\textwidth]{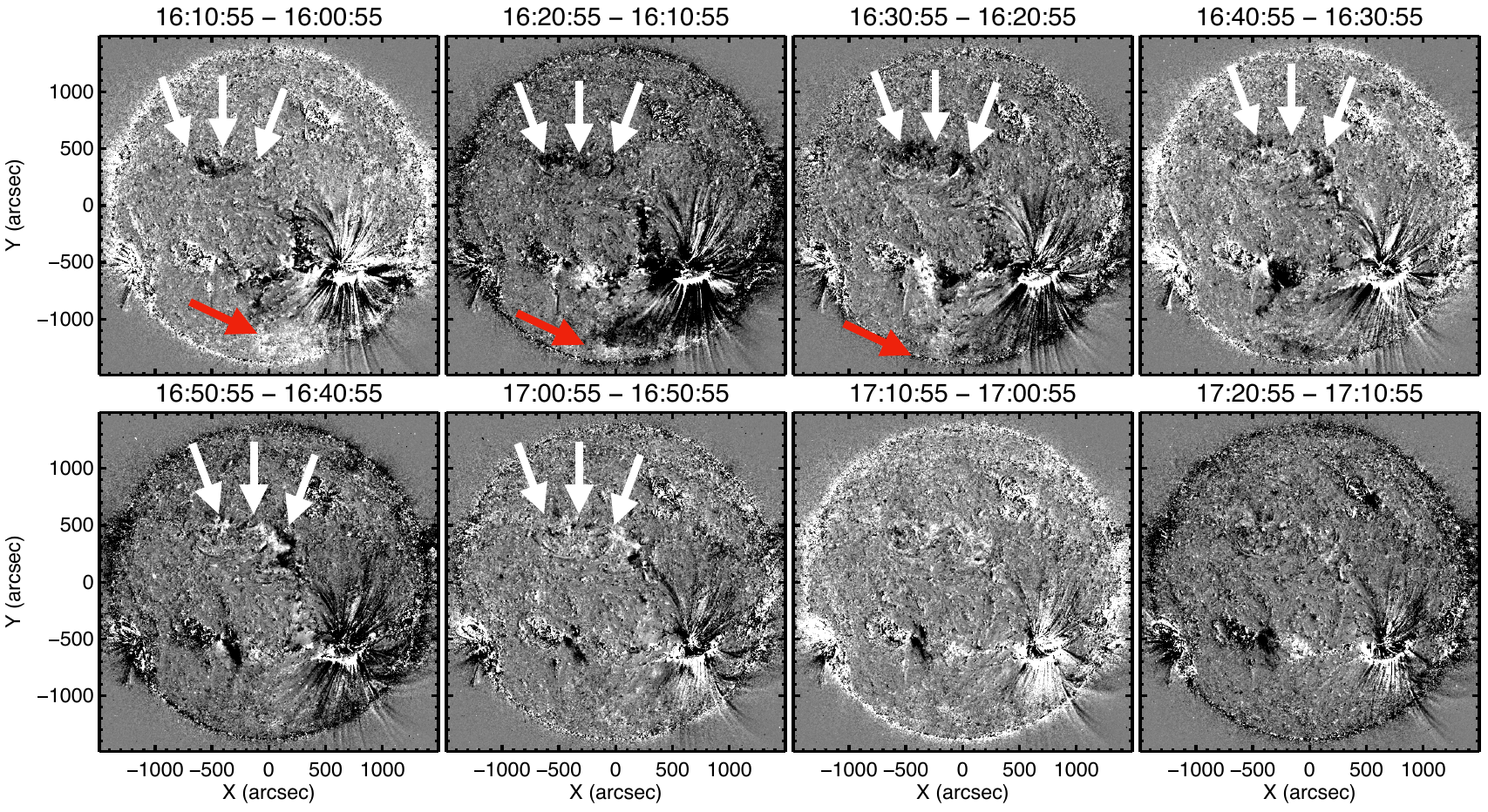}
    \caption{Running difference images showing the apparent brightenings corresponding to a supposed global EUV wave. The white arrows indicate a large dimming region which remains stationary with time, albeit with some evolution in size and shape. The red arrows indicate a feature propagating south away from the erupting active region which corresponds to a global wave well observed by the \emph{STEREO}-A spacecraft, but was behind the limb as seen from \emph{Solar Orbiter}/EUI.}
    \label{fig:wave}
\end{figure*}

\subsection{Eruption of the flux rope}
\label{ss:eruption}

Following the evolution of the bulk plasma flow along the flux rope structure, the flux rope erupted beginning at $\sim$16:00~UT on 5~September. In the low corona as observed by EUI/FSI, the eruption was associated with an apparent global EUV wave \citep[see, e.g.][for more details]{Long:2017b,Long:2017a,Long:2021}. Figure~\ref{fig:wave} shows the low coronal evolution of the eruption using a series of running difference images (produced by subtracting the image $n-1$ from the image $n$) from the 174~\AA\ passband. The white arrows in Figure~\ref{fig:wave} show a large dimming region which remains stationary with time, albeit with some evolution in size and shape. Such large-scale dimmings show that the plasma from a large part of the corona may be erupted during a CME \citep[see e.g.][]{Zhukov2007}. Although on initial inspection the evolution of this feature would appear to be the global EUV wave (the FSI image cadence here is probably too low to show the wave propagation clearly), it can be seen in Figure~\ref{fig:wave} that these arrows remain stationary with time, suggesting that rather than corresponding to a global EUV wave, this feature corresponds to the boundary of the region of influence of the erupting flux rope. In contrast, the red arrows in Figure~\ref{fig:wave} show an apparent global wave propagating south away from the erupting active region. Analysis of this event using the \emph{STEREO-A} spacecraft does suggest the presence of a global wave (A.~Vourlidas, private communication), but this propagated behind the limb as seen from \emph{Solar Orbiter}/EUI and is therefore not studied here.

\begin{figure*}[!t]
    \centering
    \includegraphics[width=\textwidth]{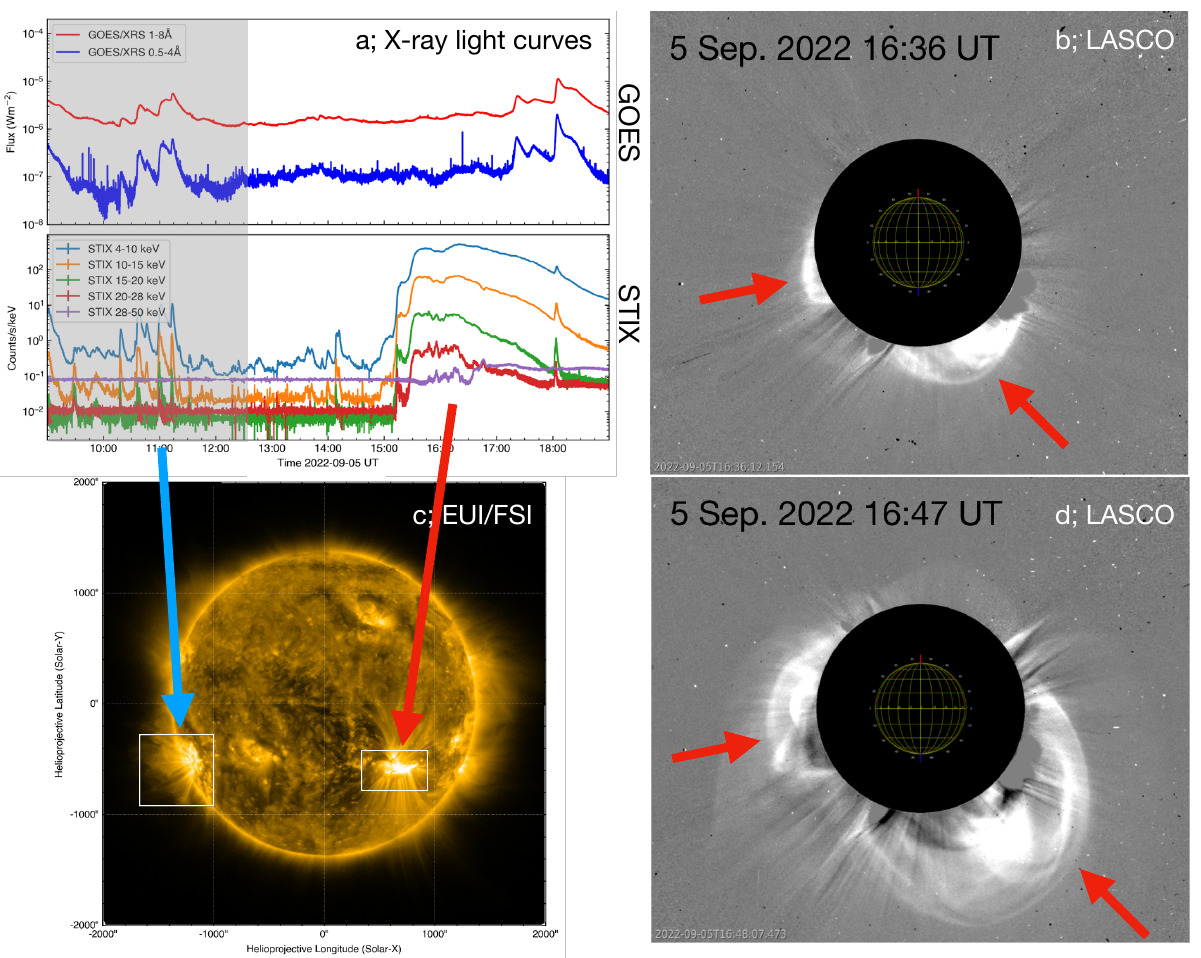}
    \caption{The flare and CME associated with the eruption on 5~September~2022. Panel~a shows the GOES (top) and STIX (bottom) X-ray light curves highlighting the flares associated with the active region rotating on-disk (grey shaded area in panel~a and blue arrow in the left panels) and the flare associated with the eruption of the flux rope (red arrow in the left panels), both of which were observed by EUI/FSI (panel~c). Panels b \& d show running difference images highlighting the erupting CME as viewed by LASCO C2. The CME is first seen in C2 at 16:36 UT and exhibits two distinct lobes as highlighted by the red arrows and discussed in the text.}
    \label{fig:eruption}
\end{figure*}

Given the location of the erupting structure on the far side of the Sun to the Earth, it was not possible to observe and quantify the associated flare using the X-ray Sensor (XRS) onboard the GOES spacecraft. Instead, the Spectrometer/Telescope for Imaging X-rays \citep[STIX;][]{Krucker:2020} onboard \emph{Solar Orbiter} was used to analyse the associated X-ray emission. It can be seen from panels a \& c of Figure~\ref{fig:eruption}, which shows X-ray lightcurves from both the GOES-XRS (a; top) and STIX (a; bottom), that while the active region traversing the limb was the primary source of X-ray emission in the lead up to the eruption (as this emission appears in both GOES-XRS and STIX lightcurves), the X-rays associated with the eruption itself must originate from the active region origin of the plasma flows described in Section~\ref{ss:brightening}, as this emission is only observed by STIX. Imaging using STIX observations at these times confirms this (although not shown here).

The eruption was initially observed in the low corona by EUI/FSI (Figure~\ref{fig:eruption}c), and subsequently as a back-sided eruption from near Earth by the Large Angle Spectroscopic Coronagraph \citep[LASCO;][]{Brueckner:1995} onboard the SOlar and Heliospheric Observatory \citep[SOHO;][]{Domingo:1995} (Figure~\ref{fig:eruption}b \& d). The CME observed by LASCO appears to have two distinct lobes as indicated by the two red arrows in panels b \& d of Figure~\ref{fig:eruption}. This appearance could be interpreted as the ends of the erupting flux rope assuming a ``croissant''-like morphology \citep[similar to that employed by the Graduated Cylindrical Shell model, e.g.,][]{Thernisien:2006, Thernisien:2011}. In contrast to the bottom-right to top-left orientation of the flux rope when on-disk as observed by \emph{Solar Orbiter}, the lobes of the brightenings observed by \emph{SOHO}/LASCO from the opposite side of the Sun suggest a much more southward directed CME. This suggests a strong deflection of the CME as it erupted. The CME also appears to be fast, advancing to cover a significant portion of the field of view of LASCO in the $\sim$10~minutes from 16:36 -- 16:47~UT.

An alternative explanation for the two distinct lobes of the erupting CME is to interpret it as two distinct CME eruptions. In this case, the extent of the large eruption to the south suggests that it is the CME associated with the eruption of the flux rope structure described here. The smaller eruption to the solar west is then more consistent with small-scale loop brightenings to the north east of the erupting active region observed at $\sim$16:00~UT (i.e., just to the right of the red arrowhead in Figure~\ref{fig:eruption}c). This would correspond to the slight increase in the STIX x-ray light-curve shown in Figure~\ref{fig:eruption}a at this time.

\subsection{\emph{In-situ} measurements}
\label{ss:insitu}

\begin{figure*}[!t]
    \centering
    \includegraphics[width=\textwidth]{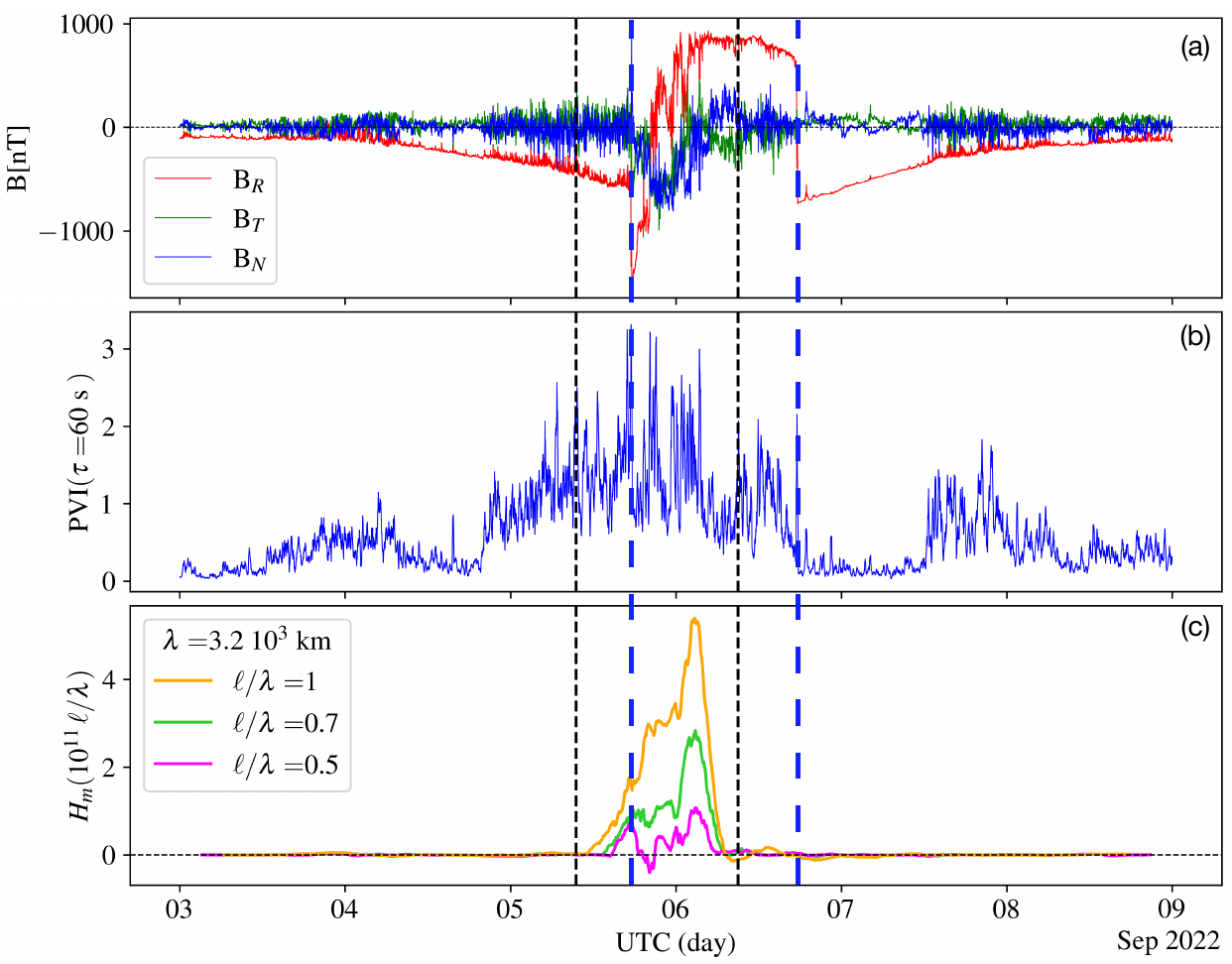}
    \caption{Examination of the flux rope detected \emph{in-situ} by \emph{Parker Solar Probe} using a characteristic length scale $\lambda=3.2 \times 10^7$~km (corresponding to a correlation time $T_{c}\sim6.4$~hrs with a nominal speed of $1000$~km/s). Stacked plots show (a) the magnetic field measured by the FIELDS instrument, (b) the PVI signal computed using a time lag of 60~s, and (c) the magnetic helicity at different scales. Black dashed vertical lines indicate the start and end of the identified flux rope structures, blue dashed vertical lines indicate discontinuities in the heliospheric magnetic field direction.}
    \label{fig:flux_rope}
\end{figure*}

As previously noted, the eruption of the flux rope coincided with Encounter 13 of PSP, as it approached a perihelion of 0.062~au on 6~September~2022. As a result of its proximity and position with respect to the Sun, PSP passed through the erupting CME, enabling a validation of the flux rope properties predicted by the remote sensing observations from \emph{Solar Orbiter}, using direct \emph{in-situ} measurements. 

Identification of the flux rope in the \emph{in-situ} data was performed using the magnetic helicity--partial variance of increments ($H_m$-PVI) technique described in \citet{Pecora:2021}. As described by \citet{matthaeus1982measurement}, the magnetic helicity at a certain scale $\ell$ can be estimated using the non-diagonal terms of the fluctuating magnetic field autocorrelation tensor, 
\begin{eqnarray}
    R_{ij}(r)&=&\langle B_i(x+r)B_j(x) \rangle \\
    \textrm{as  } H_m(x,\ell) &=& - \int_0^\ell dr_i \; \epsilon_{ijk} R_{jk},
\end{eqnarray}
where $\langle \dots \rangle$ indicates an average over a suitable interval, and $r$ describes the increments along the direction $i$. For spacecraft measurements, the increments are intended to be taken in the time domain and can be converted into spatial distances using the Taylor hypothesis \citep{taylor1938spectrum}. As usually the relative motion of the solar wind with respect to PSP is mostly radial, the direction of the increments using this technique can therefore be considered to be along the coordinate R of the RTN (Radial-Tangential-Normal) reference frame. The two transverse directions $j$ and $k$ are associated with the T and N coordinates. This quantity gives local large-scale information on helical magnetic field lines but is less sensitive to small-scale features. The technique is generally supported by the PVI \citep{Greco:2008, Greco:2018} that is sensitive to small-scale gradients and discontinuities, therefore granting a more precise detection of the boundaries of a flux tube \citep[cf.][]{pecora2019single_GSPVI,pecora2021parker_HmPVIEp}.

For this analysis, we used magnetic field measurements from 3--8~September using the FIELDS experiment \citep{Bale:2016} at a 1-minute cadence, as shown in Figure~\ref{fig:flux_rope}a. The magnetic helicity here is computed at three different characteristic scales, namely 1, 0.7, and 0.5 correlation lengths as shown in panel~c. The $H_m$-PVI technique identifies the helical structure of the CME between $\sim$09:00~UT on 5~September to $\sim$09:00~UT on 6~September, corresponding to the vertical black dashed lines, with the smaller scale profiles (at 0.7 and 0.5 $\ell/\lambda$) indicating some internal substructures. However, the complexity of the observations can be seen by the fact that the radial magnetic field does not appear to exhibit a strong change until the discontinuity delineated by the first dashed blue line. Nonetheless, the positive helicity identified by the $H_m$-PVI technique is consistent with the rotation of each of the radial, transverse, and normal magnetic field from negative to positive \citep[cf.][]{Bothmer:1998}. This behaviour is compatible with a right-handed flux rope orientation that matches the plasma flow observed using EUI/FSI and the inversion line of the photospheric magnetic field observed by PHI/FDT (see Figure~\ref{fig:SolO_context}).

In this particular case, the PVI defined as suitably normalized magnetic field vector increments,
\begin{equation}
    \Delta \mathbf{B} = \mathbf{B}(t+\tau)-\mathbf{B}(t),
\end{equation}
evaluated for a certain lag $\tau$, namely 
\begin{equation}
    \mbox{PVI}(\tau) = \frac{ |\Delta \mathbf{B}| }{\langle |\Delta \mathbf{B}|^2 \rangle}, 
\end{equation}
and depicted in panel~b for $\tau=60$~s, shows an unexpectedly long bursty region, in contrast to the usually clustered patches \citep[e.g.,][]{Greco:2018, chhiber2020clustering} and is therefore less reliable for the determination of the CME boundaries. This can be due to the superposition of several effects including the sampling of a fragmented Alfvèn zone as PSP spans heliodistances between 13 and 22 $R_\odot$ \citep{chhiber2022extended}, and the encounter with the CME leading shock and sheath region \citep[cf.][]{Davies:2021}\corr{, where the radial and transverse velocity of the solar wind with respect to PSP become comparable}. In addition, the blue dashed vertical lines in Figure~\ref{fig:flux_rope} identify discontinuities in the radial magnetic field, which complicate analysis and in the case of the second discontinuity, could indicate crossings of the heliospheric current sheet. This would suggest that PSP simultaneously passed through both the erupting magnetic flux rope and the heliospheric current sheet several times, complicating a detailed separation of these phenomena. \corr{A thorough analysis of the PSP \emph{in-situ} observations of this event can be found in Romero et al. (2023, \emph{submitted})}

\begin{figure*}[!t]
    \centering
    \includegraphics[width=\textwidth]{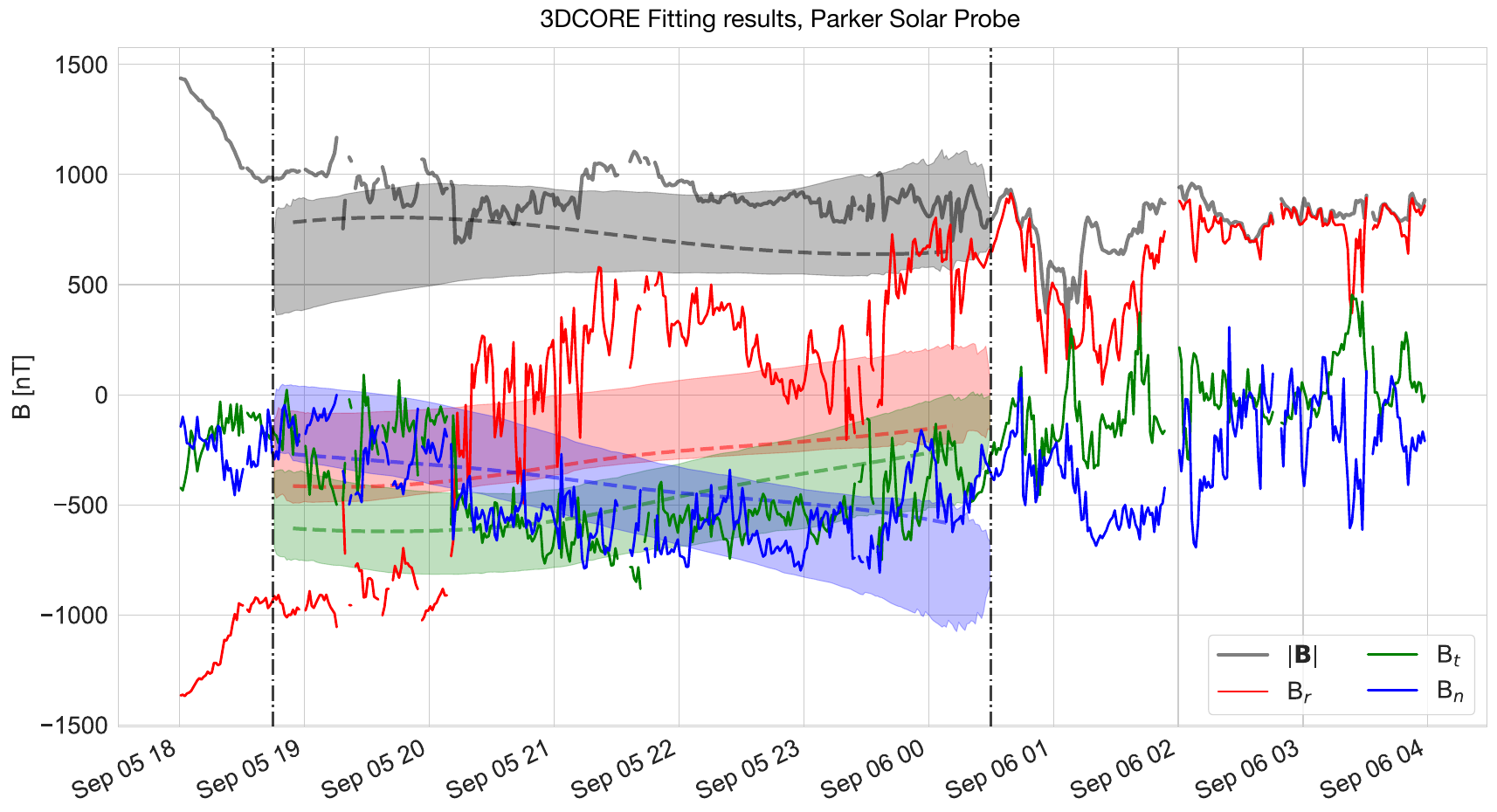}
    \caption{3DCORE model fit to the PSP/FIELDS magnetic field measurements assuming an elliptical flux rope cross-section. The dashed \corr{coloured} lines show a specific fit from the ensemble run, with the shaded areas corresponding to the $2\sigma$ spread of the ensemble. \corr{The vertical dashed black lines show the start and the end of the flux rope as taken by 3DCORE.}}
    \label{fig:3dcore}
\end{figure*}

The difficulty in determining the presence of a flux rope is highlighted by the fit produced by the 3DCORE model \citep{Mostl:2018} shown in Figure~\ref{fig:3dcore}. The 3DCORE method allows for fitting of rotating magnetic field signatures in ICMEs, assuming a Gold-Hoyle-like flux rope with an elliptical cross-section \citep{weiss2021ApJS, weiss2021AA}. The self-similarly expanding tapered torus is attached to the Sun at all times. From the fitting results we can get estimates of general flux rope parameters\corr{, including the orientation of the flux rope ($38.34^{\circ}\pm9.15$) counter-clockwise to the ecliptic plane \citep[cf.][]{Mostl:2018}}. In this study, we applied 3DCORE to the observations made using the FIELDS instrument at PSP. It is clear that while the ensemble run produced by 3DCORE does a good job of fitting the general trend of the \emph{in-situ} magnetic field evolution, there is significant variation suggesting additional effects. \corr{In addition to the issues noted in relation to Figure~\ref{fig:flux_rope} above, it should be noted that 3DCORE assumes a stationary spacecraft when fitting the detected flux rope. While this assumption works well at larger distances from the Sun, in this case PSP is moving very quickly relative to the motion of the flux rope over the spacecraft, potentially complicating the fitting of the flux rope.}

\section{Discussion}\label{s:disc} 

% Recap the observations
This study focuses on the plasma flows along, and eventual eruption of, a filament channel extending from NOAA AR~13088 across a significant portion of the solar disk into the quiet Sun. Whereas the active region portion of the filament channel was observed to contain filament plasma, the quiet sun portion was observed as a dark channel in EUV with no clear evidence of the presence of cool plasma. However, two distinct plasma flows were observed by EUI/FSI within a period of $\sim$16~hours originating in the active region and propagating along the filament channel towards the quiet Sun. The second plasma flow was then followed within $\sim$5 hours by the eruption of the filament channel. Both plasma flows exhibited distinct right-handed helical motion \citep[cf.][]{Joshi:2014}, with the first appearing to complete a complete rotation about the axis of the structure. Although the observed plasma flows occurred on 4 and 5 September~2022, the pre-eruptive evolution of the western end of the extended filament channel began with significant flux emergence into NOAA AR~13088 on 24~August and continued through to 4~September as the active region rotated over the limb as seen from Earth and onto the solar disk as observed by \emph{Solar Orbiter}. Small-scale flux cancellation could be observed along the inversion line underneath the axis of the quiet sun section of the filament channel using the very high cadence observations provided by SDO/HMI earlier in this time period, consistent with the development of a magnetic flux rope \citep[cf.][]{vanball:1989,Aulanier:1998,Yardley:2019}.

% General note about flows and filament channel structure
Deducing information about the magnetic field structure of a filament channel is typically challenging due to the absence of filamentary material within it. As a result, there is generally insufficient plasma density and consequently low EUV emission tracing out the magnetic field and providing clues as to its configuration and evolution. There have been some observations of structure within filament channels being traced out by surges and counter-streaming \corr{plasma} flows \corr{as a filament activates. These studies reveal helical field structures indicating that a flux rope may be forming or already be present} \citep[e.g.,][]{Li:2013,Yardley:2019}, but these \corr{observations} remain rare. However, it is known that filament channels contain predominantly horizontal magnetic field that is highly non-potential (i.e., aligned with the polarity inversion line) and that the \corr{channels} extend from the chromosphere into the corona \citep{Mackay:2010}. The magnetic configuration may therefore be that of a highly sheared arcade, a flux rope, or a hybrid of both. \corr{In the case in which a flux rope is inferred to be present, determining the number of turns that field lines make can be challenging unless emitting and absorbing plasma threads are present, which trace out the entirety of any helical field lines.}

As shown in previous works, filaments and filament channels may not necessarily be composed of uniformly twisted field lines that extend from one end of the channel to the other, but instead can be composed of sections with differing axial and poloidal flux values \citep{Yardley:2019}. This is especially likely to be the case in the configuration studied here as it is composed of sections located in both the weak field of the quiet sun and strong active region field, yet develops and remains stable for a period of time. Reconnection between adjacent sections of the overall configuration can then create extended magnetic field lines that enable the release of plasma from one section of the overall structure to another. However, the origin and release mechanism of that plasma flow are open questions. Previous work has shown that siphon flows \citep[e.g.,][]{Cargill:1980,Wallace:2010,Bethge:2012} can occur along magnetic flux tubes that have differing pressures at the opposite polarity footpoints. The total internal pressure of a flux tube is the sum of the magnetic and plasma pressure. As a result, a footpoint rooted in a weak magnetic field region (e.g., the quiet Sun) will have a higher plasma pressure than a footpoint rooted in a strong field region (e.g., an active region), resulting in siphon plasma flows which flow down into the footpoint in the strong magnetic field region. Other physical processes which can produce large-scale plasma flows into/within a filament channel include surges driven by magnetic reconnection near a filament channel footpoint \citep[see, e.g.,][]{Zirin:1976,Liu:2005,Chae:2003}. In this scenario, surges driven by magnetic reconnection can inject cool plasma into a pre-existing ``empty'' filament channel \citep{Liu:2005}, with that plasma potentially pooling and forming a filament or alternatively destabilising the structure and leading to its eruption.

The observations presented here of two distinct plasma flows from the leg of the filament channel rooted in the active region towards the leg rooted in the quiet Sun are at odds with the siphon flow scenario. However, they are comparable to the surge-driven injection scenario, and suggest that magnetic reconnection in the active region could have produced the observed plasma flows. The STIX observations of X-ray flux (shown in Figure~\ref{fig:eruption}a) suggest that there was no obvious X-ray emission associated with either of the plasma flows. This lack of associated X-ray emission is not unusual for surge/jet eruptions \citep[e.g.,][]{Long:2023}, but any signal may also have been masked by the increased activity from the active region rotating around the east limb as seen by \emph{Solar Orbiter} (see e.g., Figure~\ref{fig:eruption}c). The magnetogram observations provided by PHI (Figure~\ref{fig:SolO_context}a and Figure~\ref{fig:magnetograms}b) show a clear internal polarity inversion line within the active region containing the western footpoint of the filament channel, where magnetic reconnection could drive the observed plasma flows. The inset of Figure~\ref{fig:SolO_context}b also shows that a small active region filament has formed along this inversion line, which could provide a reservoir of plasma for injection into the filament channel via the observed surges. It is also interesting to note that the composition of the filament channel as observed by the SPICE spectrometer is similar to nearby quiet Sun, with no clear Mg~VIII signal in either location (see Figure~\ref{fig:spectrum}), suggesting \corr{no strong FIP effect. Previous observations have shown evidence that the quiet corona can have photospheric abundance \citep{Lanzafame:2005}, which is consistent with injection of cool filamentary plasma from the small active region filament into the flux rope observed here via a surge-driven injection.}%that coronal/chromospheric plasma has been injected into the filament channel. 

\begin{figure}[!t]
    \centering
    \includegraphics[width=0.47\textwidth]{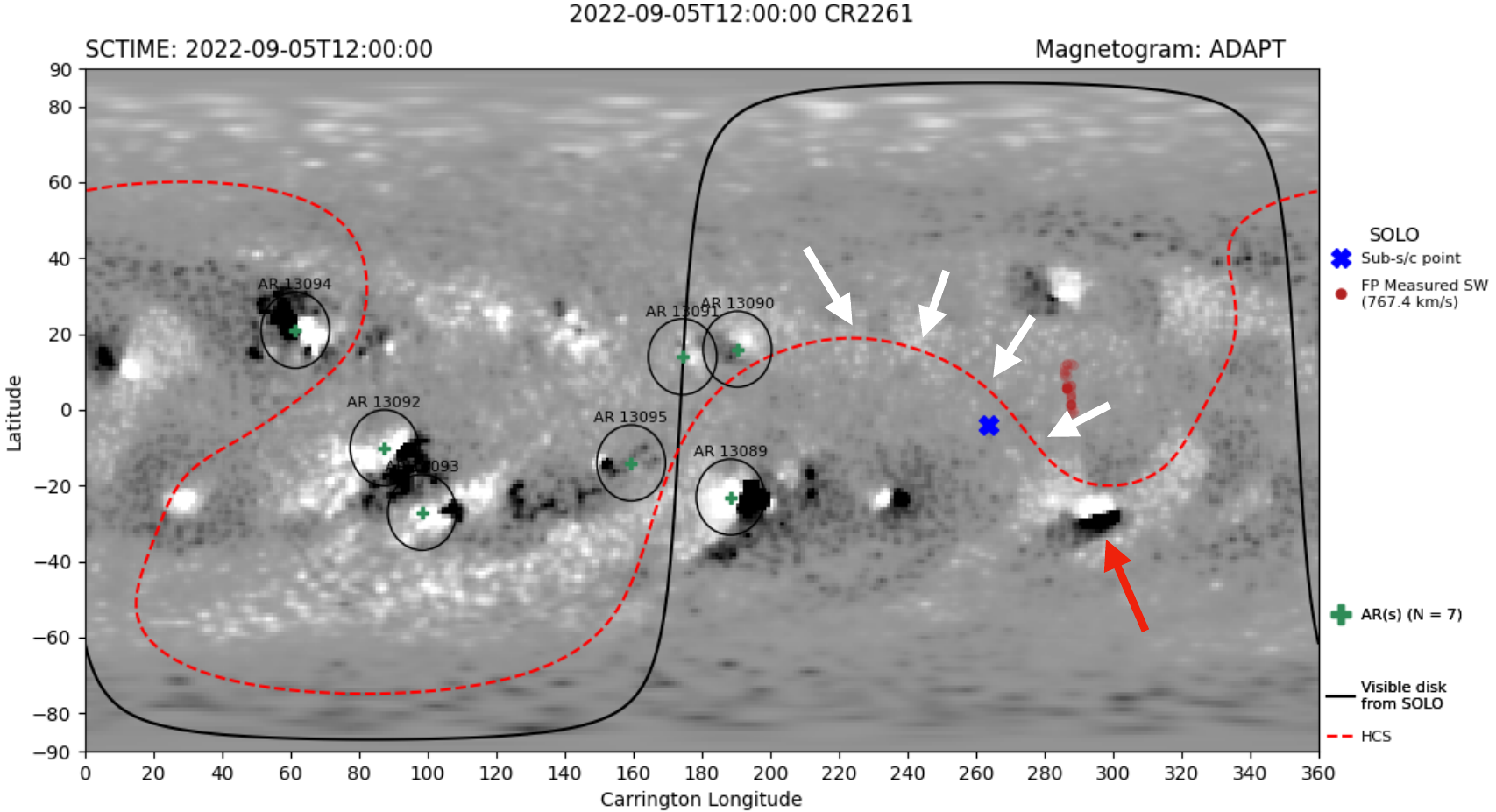}
    \caption{ADAPT magnetogram from 12:00~UT on 5~September showing the field of view (black line) and sub-spacecraft point (blue cross) of \emph{Solar Orbiter}. The red arrow indicates AR~13088 discussed here, with the white arrows indicating the location of the identified flux rope. The flux rope lies beneath the heliospheric current sheet, which could explain some of the complex structure observed in the \emph{in-situ} measurements.}
    \label{fig:hcs}
\end{figure}

The plasma flows observed by EUI/FSI, combined with the configuration of the surrounding photospheric magnetic field and the \corr{lack of observed pre-existing filamentary material}
%measured composition of the structure 
imply that this filament channel contains a magnetic flux rope which \corr{most likely formed by small-scale flux cancellation over an extended period of time \citep[\cf][]{Yardley:2019}}. However, not only do the observed plasma flows play an important role in revealing the magnetic field configuration of an otherwise very low density plasma structure, but the two distinct re-distributions of mass from the active region part of the flux rope to the quiet sun section may have had significant consequences for its stability. Previous work by \citet{Guo:2010} has shown that the additional mass and momentum imparted onto a flux rope by plasma injected into it by jets/surges can cause it to become unstable and subsequently erupt. Similarly, \citet{Seaton:2011} and \cite{Jenkins:2018} both found observational evidence of mass-unloading leading to a solar eruption, highlighting the importance of considering plasma effects as an initial driver of a solar eruption. \citet{Jenkins:2019} followed this up by developing a simple model which quantified the effect of plasma evolution in the stability of filaments\corr{. They then used this model to show that rapidly removing mass from a filament before any loss of equilibrium enabled the filament to rise sharply, noting that this effect was} more pronounced for quiescent filaments. This simple approach provided results consistent with the 3-D MHD simulations of \citet{Fan:2018} at a fraction of the computational requirements, indicating that this is a fundamental property which cannot be easily ignored. The increased length and additional kinking in the flux rope observed here for the plasma flow on 5 September compared to the plasma flow on 4 September is consistent with a slowly rising flux rope destabilised by the initial plasma flow on 4 September. The $\sim$5~hour time period between the second plasma flow and the subsequent eruption of the flux rope then suggests further destabilisation of the flux rope. However, as shown in Figure~\ref{fig:hcs}, the flux rope described here lay along the inversion line corresponding to the heliospheric current sheet. This could have delayed the eruption of the flux rope by providing additional overlying magnetic field restricting its rise. The subsequent release of energy driven by the active region at the western footpoint could then have opened this field sufficiently to erupt the whole structure. 

Following the destabilisation driven by the second observed plasma flow and the subsequent active region flare, the flux rope erupted out into the heliosphere. The footpoints and magnetic extent of the region of influence of the flux rope can be identified in the EUI/FSI observations, particularly using difference images. It is clear from Figure~\ref{fig:wave} that the flux rope encompassed a significant portion of the observed solar disk, with the coronagraph observations from LASCO exhibiting two distinct lobes (cf. Figure~\ref{fig:eruption}b \& d). Due to the proximity of the PSP spacecraft to the erupting structure, it was then possible to validate the suggestion that this was a flux rope structure using \emph{in-situ} measurements taken much closer to the Sun than ever before. 

The magnetic helicity-partial variance of increments ($H_m$-PVI) approach was used to identify the existence of a magnetic flux rope using data from the PSP/FIELDS instrument. As shown in Figure~\ref{fig:flux_rope}, this technique strongly suggests the presence of a flux rope, but instead of the clustered patches typically observed using the PVI approach, a long, bursty region was found around the observed flux rope. As noted in Section~\ref{ss:insitu}, this could be due to PSP sampling a fragmented Alfv\'{e}n zone or crossing the heliospheric current sheet. Given that the flux rope appears to have formed beneath the heliospheric current sheet (as shown in Figure~\ref{fig:hcs}), the very complex \emph{in-situ} observations shown in Figures~\ref{fig:flux_rope} and \ref{fig:3dcore} can be explained by the flux rope erupting through the heliospheric current sheet. The complexity of the scenario is then increased by the proximity of PSP to the eruption site \corr{and the resulting speed of the spacecraft}, so that the signatures have not yet been smoothed out by ambient solar wind processes.

\section{Conclusions}
\label{s:conc}

The event described here offers a unique opportunity to combine remote-sensing and \emph{in-situ} observations of the Sun from closer than ever before to gain an insight into the formation and eruption of a magnetic flux rope in the solar corona. The flux rope described here is a large intermediate flux rope which develops in the corona and is ultimately filled and destabilised by consecutive surge-driven plasma flows originating from the active region footpoint. The unique perspective of \emph{Solar Orbiter} enables a detailed analysis of this long-lived structure despite the relatively low observing cadence and switch-off of all onboard instruments due to a gravity assist manoeuvre at Venus. Following the destabilisation and eruption of the flux rope, it passed over the \emph{Parker Solar Probe} spacecraft within 14~solar radii of the Sun. This meant that it was possible to validate the conclusions drawn from the remote-sensing observations using \emph{in-situ} measurements very close to the Sun. The \emph{in-situ} measurements are consistent with the existence of a magnetic flux rope, with additional modelling of the flux rope predicting \emph{in-situ} measurements very similar to that ultimately observed. The technique used to identify the flux rope in the \emph{in-situ} magnetic field measurements also found additional structure in the data suggesting that the spacecraft contemporaneously passed through the heliospheric current sheet, consistent with global magnetic field models which indicate that the flux rope formed along the magnetic inversion line beneath the heliospheric current sheet.

This eruption highlights the benefits of having both a spacecraft with a comprehensive instrument suite combining remote-sensing and \emph{in-situ} instruments far from the Sun-Earth line like \emph{Solar Orbiter} and contemporaneous \emph{in-situ} measurements close to the Sun, as provided by \emph{Parker Solar Probe}. The observations also show the importance of plasma flows in destabilising magnetic structures in the solar atmosphere, and highlight the importance of long-term tracking of solar features away from the Sun-Earth line.

%%%%%%%%%%%%%%%%%%%%%%%%%%%%%%%%%%%%%%%%%%%%%%%%%%%%%%%%%%%%%%%%%%%%%%%%%%%
%% Acknowledgements
%
\section{Acknowledgements}
\corr{ The authors wish to thank the anonymous referee whose suggestions helped to improve the paper.} 
Solar Orbiter is a space mission of international collaboration between ESA and NASA, operated by ESA. The EUI instrument was built by CSL, IAS, MPS, MSSL/UCL, PMOD/WRC, ROB, LCF/IO with funding from the Belgian Federal Science Policy Office (BELSPO/PRODEX PEA 4000134088); the Centre National d’Etudes Spatiales (CNES); the UK Space Agency (UKSA); the Bundesministerium für Wirtschaft und Energie (BMWi) through the Deutsches Zentrum für Luft- und Raumfahrt (DLR); and the Swiss Space Office (SSO). 
The development of SPICE has been funded by ESA member states and ESA. It was built and is operated by a multi-national consortium of research institutes supported by their respective funding agencies: STFC RAL (UKSA, hardware lead), IAS (CNES, operations lead), GSFC (NASA), MPS (DLR), PMOD/WRC (Swiss Space Office), SwRI (NASA), UiO (Norwegian Space Agency).
The German contribution to SO/PHI is funded by the Bundesministerium für Wirtschaft und Technologie through Deutsches Zentrum für Luft- und Raumfahrt e.V. (DLR), Grants No. 50 OT 1001/1201/1901 as well as 50 OT 0801/1003/1203/1703, and by the President of the Max Planck Society (MPG). 
The Spanish contribution is funded by AEI/MCIN/10.13039/501100011033/(RTI2018-096886-C5, PID2021-125325OB-C5, PCI2022-135009-2) and ERDF “A way of making Europe”; “Center of Excellence Severo Ochoa” awards to IAA-CSIC (SEV-2017-0709, CEX2021-001131-S); and a Ramón y Cajal fellowship awarded to DOS.
The French contribution is funded by the Centre National d’Etudes Spatiales.
The FIELDS experiment on the Parker Solar Probe spacecraft was designed and developed under NASA contract NNN06AA01C.
DML is grateful to the Science Technology and Facilities Council for the award of an Ernest Rutherford Fellowship (ST/R003246/1), and to the attendees of the 23$^{rd}$ PSP SWG who helped clarify the scientific interpretation of this event. 
The work of DHB was performed under contract to the Naval Research Laboratory and was funded by the NASA Hinode program.
The work is supported by PSP HelioGI under grant number 80NSSC21K1765 at University of Delaware and by the Parker Solar Probe mission (NNN06AA01c) under a subcontract at Delaware (SUB0000165) from the ISOIS project at Princeton University.
The ROB team thanks the Belgian Federal Science Policy Office (BELSPO) for the provision of financial support in the framework of the PRODEX Programme of the European Space Agency (ESA) under contract numbers 4000134474 and 4000136424.
EED and UVA are funded by the European Union (ERC, HELIO4CAST, 101042188). Views and opinions expressed are however those of the author(s) only and do not necessarily reflect those of the European Union or the European Research Council Executive Agency. Neither the European Union nor the granting authority can be held responsible for them.
%\end{acknowledgements}

\facilities{SDO/AIA, SO EUI, SPICE, PHI, PSP/FIELDS}

\software{SSW/IDL \citep{Freeland:1998}, SunPy \citep{Sunpy:2020}, Matplotlib \citep{Hunter:2007}, AstroPy \citep{Astropy:2022}}

\bibliography{bibliography}{}
\bibliographystyle{aasjournal}

\end{document}